\documentclass[11pt]{article}
\usepackage{graphicx,ctable}
\begin{document}
\begin{center}
{\Large\bf{On the precise determination of the Tsallis parameters in proton - proton collisions at LHC energies}}\\
\vspace*{1cm}  
{\sf{T. Bhattacharyya$^1${\footnote{email: trambak.bhattacharyya@uct.ac.za}}, J. Cleymans$^1${\footnote{email: jean.cleymans@uct.ac.za}}, 
L. Marques$^2${\footnote{email: lmarques.if@gmail.com}},\\[0.1cm]
 S. Mogliacci$^1${\footnote{email: sylvain.mogliacci@uct.ac.za}}, M.~W.~Paradza$^{1,3}${\footnote{email: mwparadza@gmail.com}}}}\\
\vspace{10pt}  
{\small  {\em $^1$UCT-CERN Research Centre and Department of Physics, University of Cape Town,\\ Rondebosch 7701, South Africa\\
$^2$Instituto de F\'isica, Universidade de S\~ao Paulo-IFUSP, Rua do Mat\~ao,\\
           Travessa R187, 05508-900 S\~ao Paulo, S\~ao Paulo, Brazil\\
$^3$  Department of Physics and Astronomy,\\
     University of The Western Cape, Bellville 7535, South Africa}}
\normalsize  
\end{center}

\begin{abstract}
A detailed analysis is presented of the precise values of the Tsallis parameters obtained in $p-p$ collisions 
for identified particles, pions, kaons and protons at the LHC at three beam energies $\sqrt{s} = 0.9, 2.76$ and $7$ TeV. 
Interpolated data at $\sqrt{s} = $ 5.02 TeV have also been included.
It is shown that
the Tsallis formula provides reasonably good fits to the $p_T$ distributions in 
$p-p$ collisions at the LHC using three parameters $dN/dy$, $T$ and $q$.
However, the parameters $T$ and $q$ depend on the particle species and are different for pions, kaons and protons.
As a consequence  there is no $m_T$ scaling and also  no universality of the parameters
for different particle species.
\end{abstract}
\noindent
PACS Indices:~12.38.Aw,~13.60.Le,~14.40.Lb,~14.40Nd
\vspace{1mm}
\section{Introduction}

It is well accepted that the transverse momentum distributions in high energy $p-p$ 
collisions are described by a power law distribution at the Relativistic Heavy Ion Collider 
(RHIC)~\cite{STAR,PHENIX1,PHENIX2} as well as at the Large Hadron Collider (LHC)~\cite{ALICE2,CMS1,CMS2,ATLAS,ALICE_charged}. 

In this paper we investigate in detail one particular form of power law distribution which has been used extensively in the literature
~\cite{bialas,grigoryan,wong,ristea,biro,parvan1,parvan2,tripathy,zheng1,marques,azmi,sorin,sena,Azmi:2014dwa,Biro:2004qg,Khuntia:2017ite,Khandai:2013gva,zheng,de}
given by~\cite{tsallis}
\begin{equation}
f(E) \equiv \left[ 1 +(q-1) \frac{E}{T} \right]^{-1/(q-1)} ,
\end{equation}
where $q$ and $T$ are two parameters to be determined.
It is referred to as the Tsallis distribution~\cite{tsallis} and forms the basis for non-extensive statistical thermodynamics. 
It has been shown~\cite{worku1,worku2} that  the corresponding pressure, the   particle number  and  energy densities obey
the usual thermodynamic consistency relations and the parameter $T$ is indeed related to the entropy via
\begin{equation}
\frac{1}{T} = \left.\frac{\partial S}{\partial E}\right|_{V,N} ,
\end{equation}
albeit that the entropy $S$ is the Tsallis entropy and not the standard Boltzmann-Gibbs one. Therefore we refer 
to the parameter $T$ as temperature. The physical interpretation of the parameter $q$ and 
the connection between the Tsallis temperature and the Boltzmann-Gibbs one were elucidated in~\cite{wilk}.
A few years ago it was suggested that the parameters appearing in the Tsallis distribution are 
the same for a wide range of identified hadrons~\cite{worku1,worku2} at $\sqrt{s}$ = 900 GeV in $p-p$ collisions.
Subsequently several analyses have appeared which do not support this conclusion.
Due to the size of the errors and the uncertainties on some of the parameters it was not 
completely possible to eliminate this option at $\sqrt{s}$ = 900 GeV but further analyses at higher energies
are not in support of the original ansatz and  led to the proposal of having  
sequential freeze-outs depending on the particle type~\cite{Thakur:2016boy,la,lacey,thakur}.

In this paper we focus on the data of the ALICE collaboration~\cite{ALICE_900,ALICE_2760,ALICE_7000} 
and determine the parameters appearing in the Tsallis distribution as precisely
as possible at beam energies of $\sqrt{s}$ = 0.9, 2.76 and 7 TeV.  We have also included the interpolated 
spectra provided by the ALICE collaboration  at $\sqrt{s}$ = 5.02 TeV~\cite{ALICE_5020}.

The fits were performed over the full range of the available $p_T$ range.
In order to test the quality of the extrapolation to very high values of $p_T$ we included at the end a fit to the 
spectrum of charged particles as measured by the CMS collaboration~\cite{CMS_highpt} which extends to values of $p_T$ up to 
200 GeV.

The conclusion we reach is that for the $\pi$'s, $K$'s and protons the parameters are different and no universality in the 
parameters exists. 
Thus, even though the Tsallis distribution provides a reasonable description of the transverse momenta distributions,
it has to be concluded that the parameters are clearly different. 

In this paper we do not add radial flow to the Tsallis distribution.  
A considerable effort was spent on making fits with radial flow 
using a blast-wave formalism as was done by the authors in references~\cite{zheng,Jiang:2013gxa}
but in the end we  concluded that radial flow, even 
though it adds at least one parameter to the fitting procedure, does not improve the quality of the fits in $p-p$ collisions.  
We note that reference~\cite{grigoryan} 
 uses a similar procedure  but with
a considerable number of parameters (at least 8). Also the radial flow is made to depend on the hadron considered, it is
different for pions, kaons and protons. In the end we decided to go for the simplest form as far as possible and use
the smallest number of parameters. 
Also, our results show that the 
temperature $T$ is comparable for kaons and for protons at higher beam energies which does not have 
a natural explanation in the radial flow scenario.

One open question, concerning the effect of feed-down decays from resonances, remains. This has been
considered in great detail in  $e^+e^-$ annihilation in~\cite{ferroni}. The calculations are however much more involved and at the
moment it does not look feasible to repeat such an analysis in the energy range of the LHC.

\section{Tsallis Distribution}

The following form is used for the transverse momentum distribution of hadrons produced at LHC energies~\cite{worku1,worku2} 
\begin{equation}
\left.\frac{1}{p_T}\frac{d^{2}N}{dp_T~dy}\right|_{y=0} = gV\frac{m_T}{(2\pi)^2}\left[1+(q-1)\frac{m_T}{T}\right]^{-q/(q-1)} ,
\label{tsallisfit}
\end{equation}
where $p_T$ and $m_T$ are the transverse momentum and transverse mass respectively, $y$ being the rapidity,
$g$ is the degeneracy factor, i.e. $ g = 1$ for $\pi^+, \pi^-, K^+, K^-$ since these are pseudoscalar particles 
while $ g = 2$ for protons and 
antiprotons because of their spin 1/2 nature. $V$ is the volume of the system. 
Since particles and antiparticles are produced 
equally abundantly at LHC energies there is no need for the introduction of chemical potentials. 
At lower energies it would, of course, be necessary to take a chemical potential into account. 
The resulting values of $q$ and $T$ are for a system at kinetic freeze-out.
The fact that the right-hand side of Eq.~(\ref{tsallisfit}) only depends on $m_T$ is known as $m_T$ scaling.

Eq.~\ref{tsallisfit} is a generalisation of the standard 
Boltzmann-Gibbs distribution,  in the limit where $q -> 1$  one recovers
\begin{equation}
\lim_{q\rightarrow 1}\left.\frac{1}{p_T}\frac{d^{2}N}{dp_T~dy}\right|_{y=0} = gV\frac{m_T}{(2\pi)^2}
                          \exp\left(-\frac{m_T}{T}\right)
\label{boltzmann}
\end{equation}
hence, many concepts familiar from statistical mechanics can be carried over to the distribution presented in Eq.~\ref{tsallisfit}.

As noted in~\cite{poland}, Eq.~(\ref{tsallisfit}) can be integrated over the transverse momentum, leading to
\begin{eqnarray}
\frac{dN}{dy}\bigg|_{y=0} & = &\frac{gV}{(2\pi)^2} \int_0^\infty p_T\,dp_T\,m_T \bigg[1+(q-1)\frac{m_T}{T} \bigg]^{-\frac{q}{q-1}}\nonumber\\
 & = &\frac{gVT}{(2\pi)^2}\bigg[\frac{(2-q)m_0^2 + 2m_0T + 2T^2}{(2-q)(3-2q)} \bigg]\bigg[1+(q-1)\frac{m_0}{T} \bigg]^{-\frac{1}{q-1}} ,
\label{poland_N}
\end{eqnarray}
where $m_0$ is the mass of the considered particle.

From Eq.~(\ref{poland_N}), we can express the volume in terms of $dN/dy$ and the parameters $q$ and $T$  as
\begin{equation}
 V = \frac{dN}{dy}\bigg|_{y=0} \frac{(2\pi)^2}{gT}\bigg[\frac{(2-q)(3-2q)}{(2-q)m_0^2 + 2m_0T + 2T^2} \bigg]\bigg[1+(q-1)\frac{m_0}{T} \bigg]^{\frac{1}{q-1}} .
\end{equation}

Thus, replacing the volume in Eq.~(\ref{tsallisfit}) by the more accessible $dN/dy$ it can be rewritten as
\begin{eqnarray}
\frac{1}{p_T} \frac{d^2N}{dp_T\,dy}\bigg|_{y=0}&=& \frac{dN}{dy}\bigg|_{y=0} \frac{m_T}{T} \frac{(2-q)(3-2q)}{(2-q)m_0^2 + 2m_0T + 2T^2}\bigg[1+(q-1)\frac{m_0}{T} \bigg]^{\frac{1}{q-1}}\nonumber\\
&&\bigg[1+(q-1)\frac{m_T}{T} \bigg]^{-\frac{q}{q-1}} .\,
\label{newpt}
\end{eqnarray}

It has to be noted that, in order to have a positive number of particles, $q$ must  be less than the value 3/2. A similar
consideration for a positive energy density leads to an even stronger constraint, namely $q$ \textless 4/3~\cite{Ishihara:2016moh,mogliacci}.

\indent It is also worth mentioning that the parameterization given in Eq.~(\ref{newpt}) is close to 
the one used for fitting the data taken at RHIC and LHC experiments
~\cite{STAR,PHENIX1,PHENIX2,ALICE2,CMS1,CMS2,ATLAS,ALICE_900}, which is given by
\begin{equation}
\frac{d^2 N}{dp_T dy}\bigg|_{y=0}=p_T\frac{dN}{dy}\bigg|_{y=0}\frac{(n-1)(n-2)}{nC(nC + m_{0}(n-2))}
\left[1 + \frac{m_T - m_0}{nC}\right]^{-n},
\label{BIG}
\end{equation}
where $n, C$ and $m_0$ are fit parameters. There is no justification in calling the parameter $C$ a temperature.
Notice that Eq.~(\ref{BIG}) shows the same dependence on the transverse momentum as Eq.~(\ref{tsallisfit}) 
except for an additional factor $m_T$ which is present in Eq.~(\ref{newpt}) but not in Eq.~(\ref{BIG}).  Omission of this
additional factor of $m_T$  leads to values of $T$ which are clearly larger, about double the values presented in 
this paper.

\section{Analysis of the transverse momentum spectra}
\begin{figure}
\centering
\includegraphics[width=\textwidth, height=12.0cm]{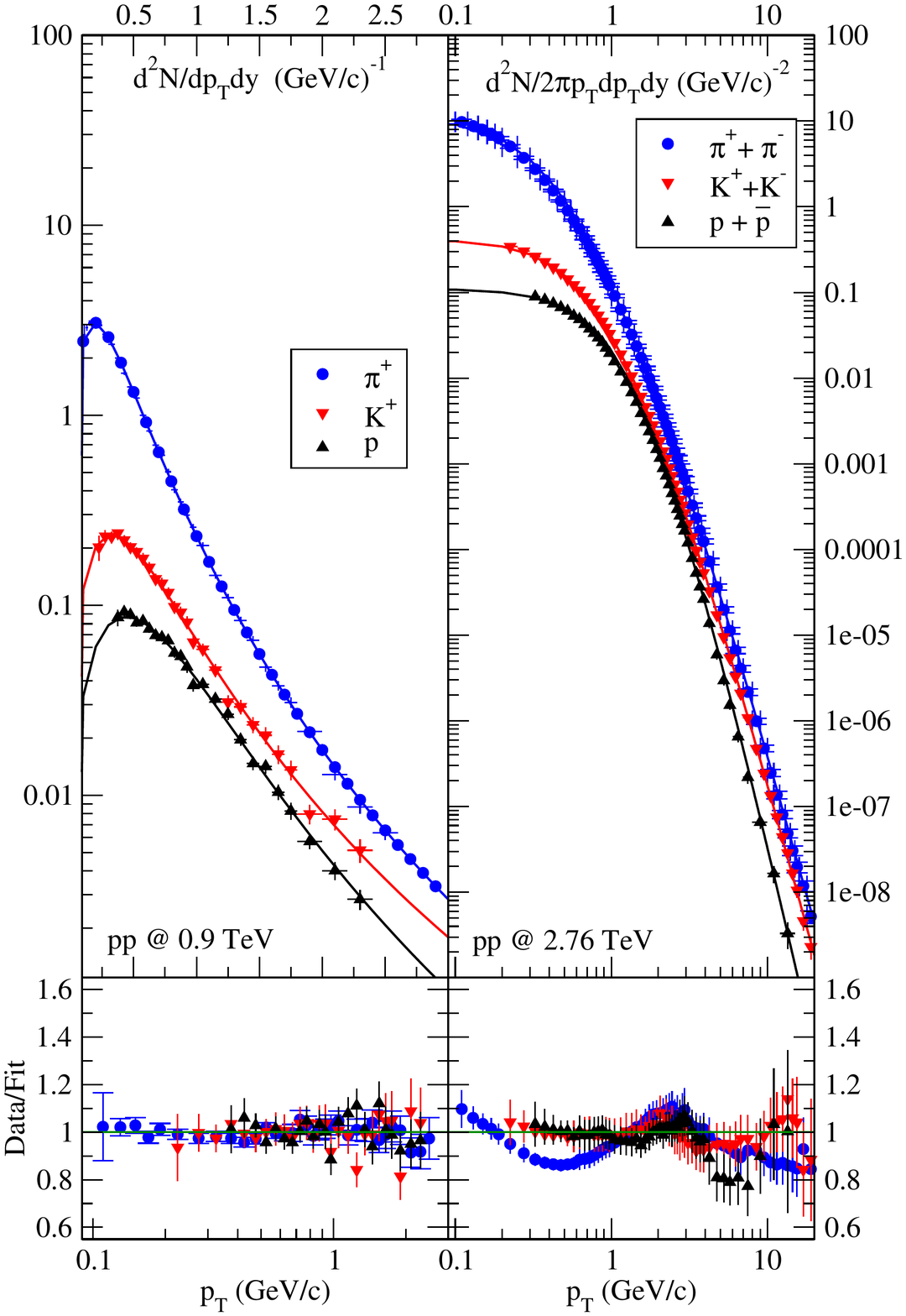}
\caption{\label{pt_low} Fits to the transverse momentum distributions, using the Tsallis distribution given by  Eq.~(\ref{newpt}),
 of $\pi^+$ (blue), $K^+$ (red) and   protons (black)   at 
900 GeV~\cite{ALICE2} in the left pane. In the right pane fits corresponding to $\pi^++\pi^-$ (blue), $K^++K^-$ (red),  protons and 
antiprotons (black)  as measured by the ALICE collaboration at 
2.76 TeV~\cite{ALICE_2760} are shown. 
        }
\end{figure}

\begin{figure}
\centering
\includegraphics[width=\textwidth, height=12.0cm]{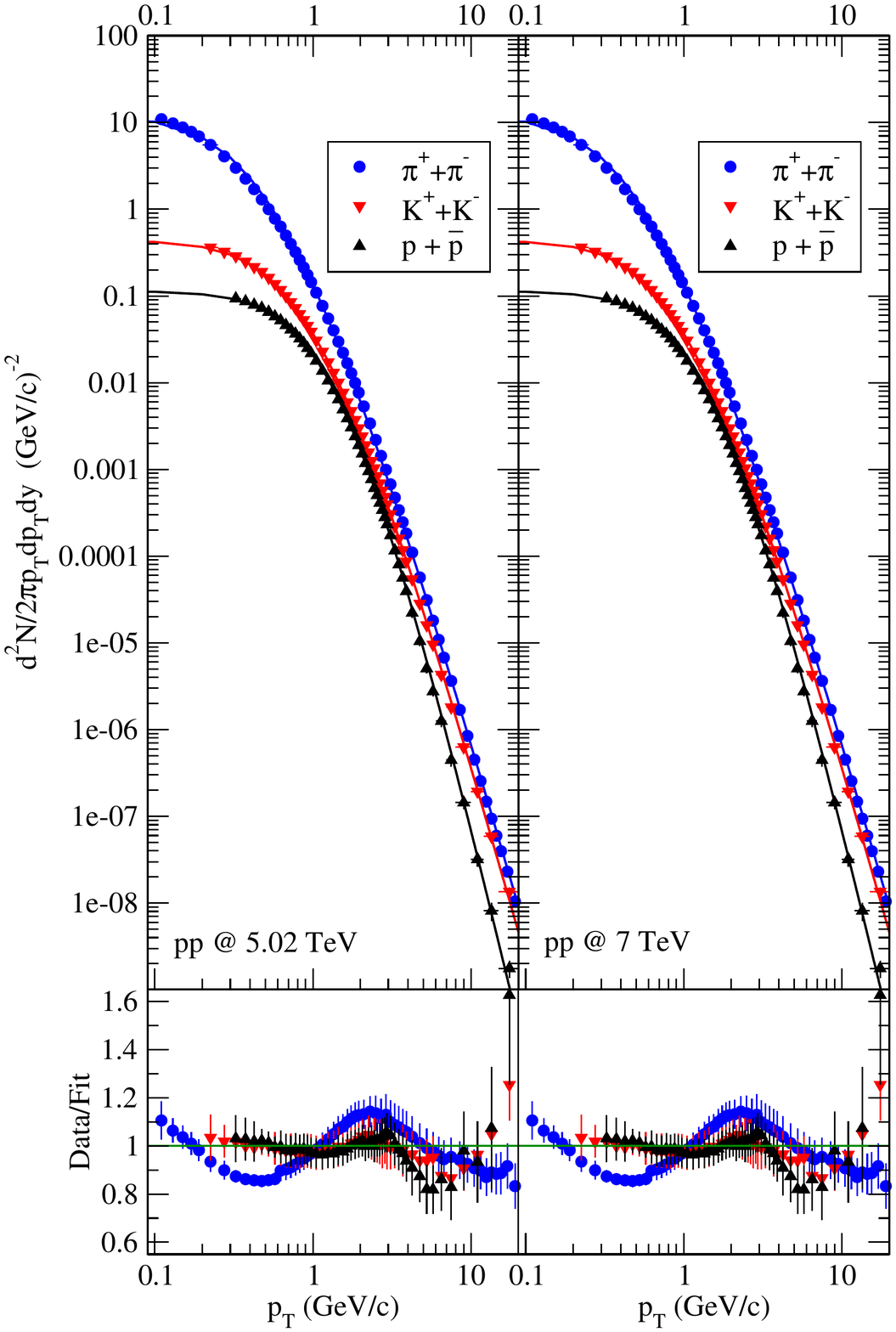}
\caption{\label{pt_high} 
Fits to the transverse momentum distributions, using the Tsallis formula given in Eq.~(\ref{newpt}), of 
$\pi^++\pi^-$ (blue), $K^++K^-$ (red),  protons and 
antiprotons (black) as given  by the ALICE collaboration at 
5.02 TeV~\cite{ALICE_5020}, by interpolating between the measured  7 TeV~\cite{ALICE_7000} (shown in the right pane) data 
and the measured 2.76 TeV data.}
\end{figure}

The Tsallis parameters were determined by fitting the experimental results published by the 
ALICE collaboration for $p-p$ collisions at three different beam energies~\cite{ALICE_900, ALICE_2760,ALICE_7000}
to Eq.~(\ref{newpt}).  
For completeness we also considered the data extrapolated by the ALICE collaboration at $\sqrt{s} = $ 5.02 TeV~\cite{ALICE_5020}.  
The results are shown in Figures~\ref{pt_low} and \ref{pt_high} 
and will be discussed sequentially in the following subsections. 

\subsection{Fits to transverse momentum spectra at the beam energy of 900 GeV}

Of all the beam energies considered in this analysis, the one at 900 GeV has the 
smallest range in $p_T$, namely about an order of magnitude less than the ones at 2.76 and 7 TeV.
Figure~\ref{pt_low} shows the results for particles $\pi^+, K^+$ and $p$. 
The results for $\pi^-$ and $K^-$ are compatible with those for $\pi^+$ and $K^+$ and are not shown.

\begin{figure}
\centering
\includegraphics[width=0.8\textwidth, height=12.0cm]{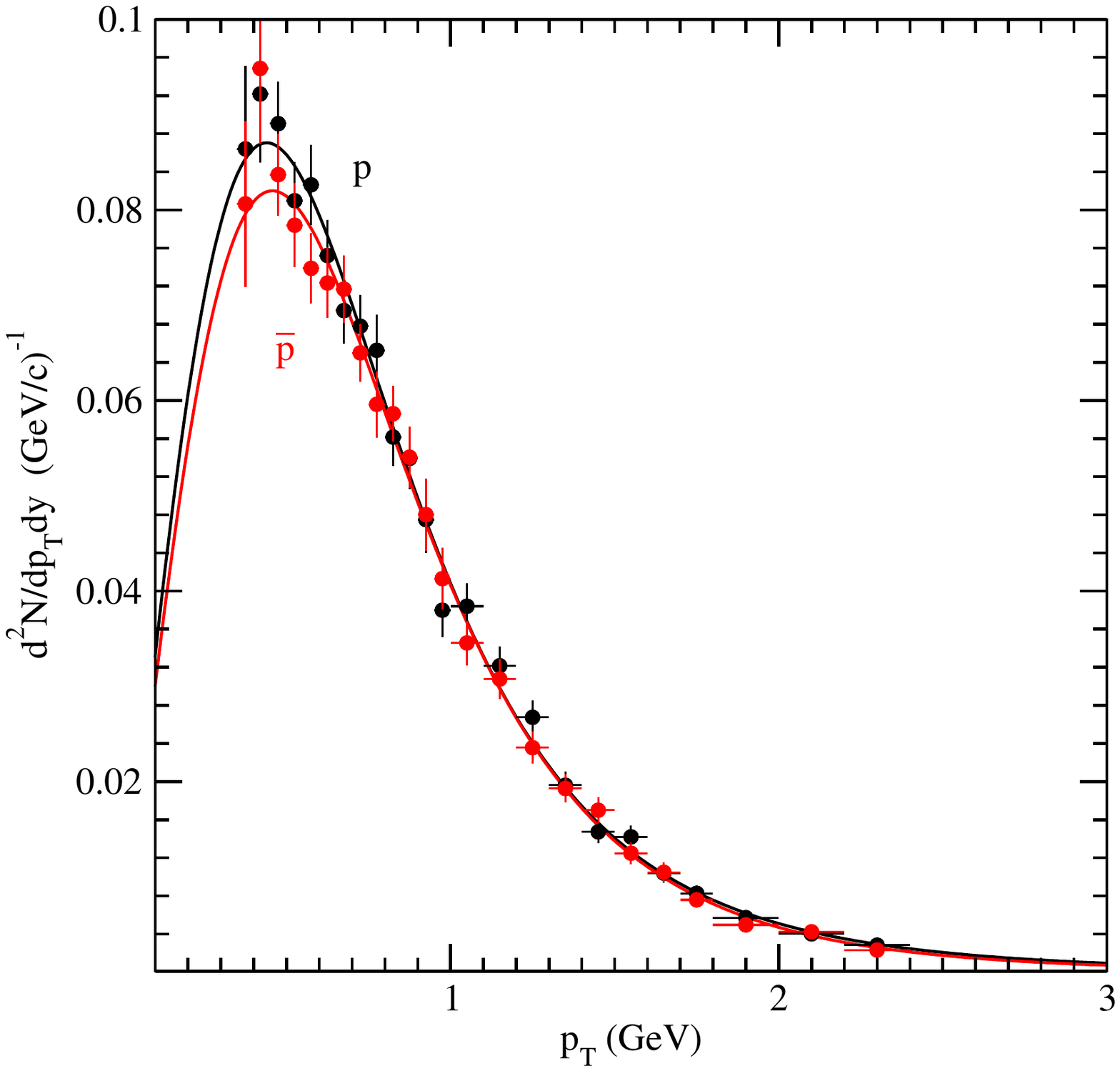}
\caption{\label{ptdistro_proton_antiproton}
Comparison of the proton and antiproton distributions in transverse momentum for $\sqrt{s}$ = 900 GeV. 
Values for protons are  in black while those for antiprotons are given in red. 
The large uncertainty  has its origin in the differences at small $p_T$.
}
\end{figure}

The comparison between protons and antiprotons is shown separately in Fig.~\ref{ptdistro_proton_antiproton}
and deserves some further discussion.
As can be seen, the difference comes mainly from the (low) $p_T < $ 0.6 GeV region, while the two distributions 
nicely overlap for higher 
values of $p_T$. It is to be noted that the integrated quantities, i.e. the values of $dN/dy$ as well as $<p_T>$,  are in 
good agreement.
The resulting parameters have large 
errors but are compatible within one standard deviation: $T$ for protons is 17 $\pm$ 28 MeV with $q$ being 1.160$\pm$ 0.025; 
for antiprotons the value of $T$ is 52 $\pm$ 30 while $q$ is 1.132 $\pm$ 0.025.  A combined fit of protons and antiprotons yields 
values with smaller errors, namely  for $T$ = 47 $\pm$ 10 MeV and  $q$ = 1.136 $\pm$ 0.007. 
The central values of $T$ thus differ by a factor of more than two between protons and antiprotons.

A summary of the resulting values for the  beam energy of $\sqrt{s} = 900$ GeV is shown in Table~\ref{tab:results_900}. 
Especially, the values 
for $T$ deviate clearly from those obtained for $\pi^{\pm}$ and $K^{\pm}$ even for the combined fit of protons and antiprotons. 
An earlier analysis gave 
consistent values, however, with large error bars~\cite{worku1}, these are not confirmed by the present analysis.

\begin{table}[!ht]
\centering
\begin{tabular}{|c|c|c|c|c|c|}\toprule
Particle     & $dN/dy$             & $q$                   & $T$ (GeV)      & $\chi_y^2$ / NDF          & $\langle p_T \rangle$ (GeV)\\ \midrule

$\pi^+$   & 1.488 $\pm$ 0.019      & 1.148 $\pm$ 0.005     & 0.070 $\pm$ 0.002    & 22.73 / 30  & 0.408 $\pm$ 0.023 \\

$\pi^-$   & 1.479 $\pm$ 0.018     & 1.145 $\pm$ 0.005     & 0.072 $\pm$ 0.002    & 15.83 / 30  & 0.408 $\pm$ 0.022 \\

$K^+$     & 0.184 $\pm$ 0.004   & 1.175 $\pm$ 0.016     & 0.057 $\pm$ 0.012    & 13.02 / 24   & 0.663 $\pm$ 0.142 \\

$K^-$     & 0.182 $\pm$ 0.004    & 1.161 $\pm$ 0.015     & 0.064 $\pm$ 0.012    & 6.214 / 24   & 0.641 $\pm$ 0.202 \\

$p$       & 0.083 $\pm$ 0.002     & 1.160 $\pm$ 0.025     & 0.017 $\pm$ 0.028    & 14.52 / 21  &  0.773 $\pm$ 0.270\\

$\bar{p}$ & 0.079 $\pm$ 0.002   & 1.132 $\pm$ 0.025    & 0.052 $\pm$ 0.030     & 13.82 / 21 & 0.766 $\pm$ 0.250\\
$p$ + $\bar{p}$ &    & 1.136 $\pm$ 0.007    & 0.047 $\pm$ 0.010    &  & \\
\bottomrule
\end{tabular}
\caption{Fit results at $\sqrt{s} $ = 900 GeV, using data 
from the ALICE~\cite{ALICE2} collaboration. Note the very large errors on the values of $T$ for protons and antiprotons.
The last line shows the results of a combined fit to the proton and antiproton distributions which leads to smaller errors.
}
\label{tab:results_900}
\end{table}

\begin{figure}
\centering
\includegraphics[width=0.8\textwidth, height=12.0cm]{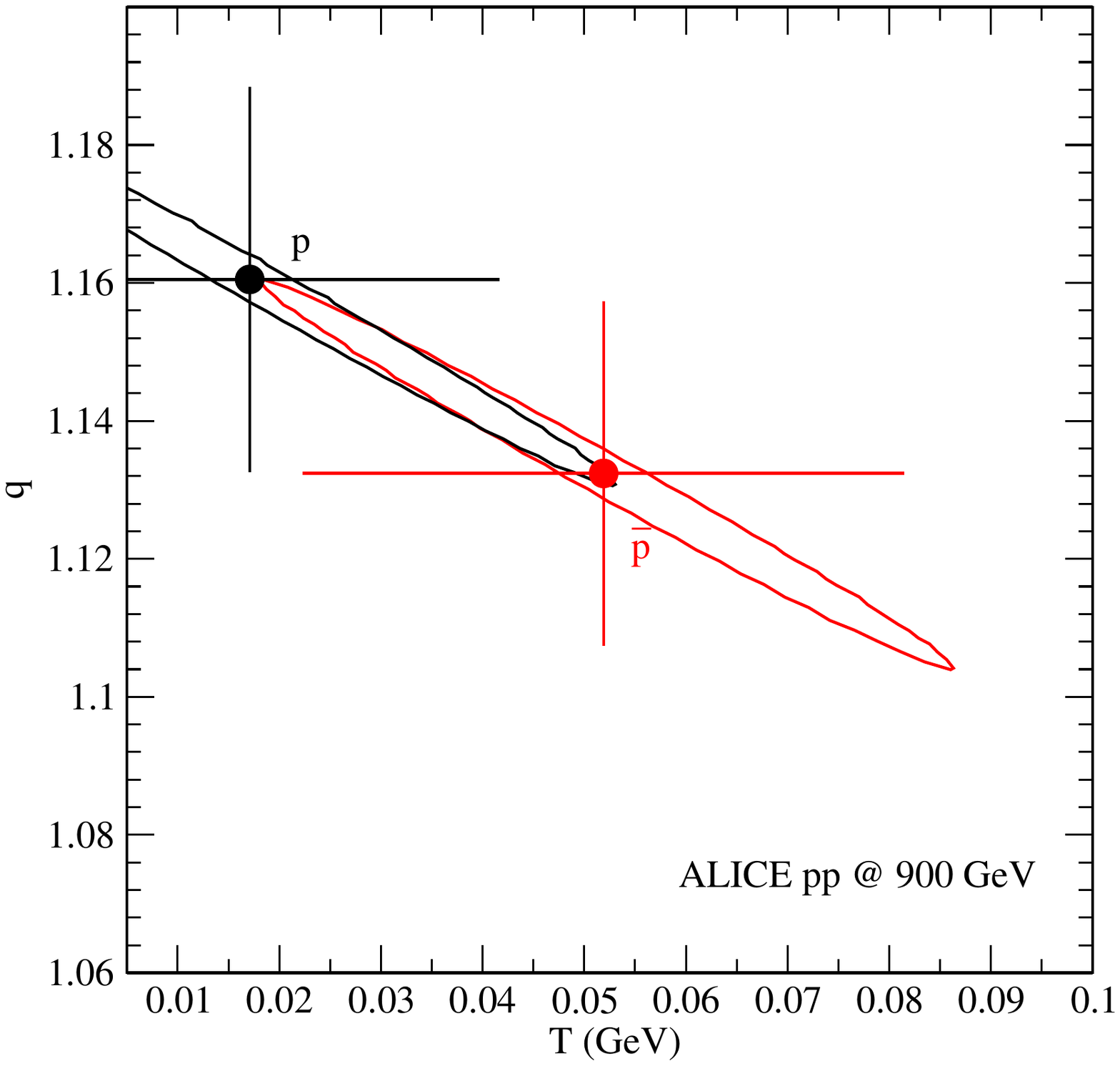}
\caption{\label{proton_antiproton}
Contours in the $T - q$ plane showing lines with 1-$\sigma$ deviation from the minimum $\chi^2$ values as well as the values of 
the parameters $q$ and $T$ (corresponding to the minimum $\chi^2$)
with error bars.
Values for protons are  in black while those for antiprotons are given in red. 
Note the large difference of more than a factor of two between the central temperature values.  
       }
\end{figure}

This difference in the  Tsallis parameters comes as a surprise since the integrated proton and antiproton 
yields are equal within the uncertainties. Our interpretation is that within the measured range of 
transverse momenta, it is not possible to determine precisely the values of the Tsallis parameters. 
To emphasize this we show in Figure~\ref{proton_antiproton} the contour of 1-$\sigma$ uncertainties in 
the $T - q$ plane. 

The proposed universality of the Tsallis parameters at $\sqrt{s}$ = 900 GeV can therefore not be established. 
There is no ambiguity in determining the parameters for the pion distributions. 
Differences are apparent at very small values of $p_T$  as can be clearly seen in Figure 3.

The values of  $\chi^2$ listed in the Table~\ref{tab:results_900} were determined as follows. 
Using the values of the transverse momentum distribution for a given value of $p_T$ 
\begin{equation}
N_i(T,q, dN/dy) \equiv  \left.\frac{d^2N}{dp_T dy}\right|_{p_T = p_T^i}
\end{equation}
for the data at 900 GeV and by 
\begin{equation}
N_i(T,q, dN/dy) \equiv  \frac{1}{2\pi p_T} \left.\frac{d^2N}{dp_T dy}\right|_{p_T = p_T^i}.
\end{equation}
for all other beam energies.
Both are  calculated using the Tsallis distribution
 with the optimised values of $q$, $T$ and $dN/dy$. 
We then compare this theoretical 
result with the experimental values to determine a value of $\chi^2$
\begin{equation}
\chi_y^2 = \sum_{i=1}^N\left[\frac{N_i(T,q,dN/dy) -  E_i}{\sigma_i}\right]^2 
\end{equation}

where  $E_i$ is the experimental value of the momentum distribution at the same value of $p_T$ and 
$\sigma_i$ is the experimental error on the distribution.  A similar procedure was used in all 
the subsequent tables. We add the index $y$ to emphasize that these values  refer to the error bars on the ordinate ($y$) axis.

\subsection{Fits to transverse momentum spectra at the beam energy of 2.76 TeV}

The transverse momentum spectra at $\sqrt{s}$ = 2.76 TeV have been measured in a range extending up to about 20 GeV/c.
The fit is shown in Figure~\ref{pt_high} and the resulting values of the parameters are listed in Table 2.
It can be seen that the values of $q$ and $T$ are much more constrained than in the previous case of $\sqrt{s}$ = 900 GeV.
If the possibility of common values could not be entirely excluded in the previous case, there is no doubt here that the
values are different.

The resulting values for the  beam energy of $\sqrt{s} = 2.76$ TeV are shown in Table~\ref{tab:results_2760}.

\begin{table}[!ht]
\centering
\begin{tabular}{|c|c|c|c|c|c|}\toprule
Particle       & $dN/dy$           & $q$                 & $T$ (GeV)           & $\chi_y^2$ / NDF & $\langle p_T \rangle$ (GeV)\\ \midrule

$\pi^++\pi^-$  & 3.967 $\pm$ 0.083 & 1.149 $\pm$ 0.002 & 0.077 $\pm$ 0.001 & 242.8 / 60 & 0.442 $\pm$ 0.011 \\

$K^++K^-$      & 0.463 $\pm$ 0.010 & 1.144 $\pm$ 0.002 & 0.096 $\pm$ 0.003 & 10.55 / 55  & 0.706 $\pm$ 0.024 \\


$p+\bar{p}$    & 0.209 $\pm$ 0.006 & 1.120 $\pm$ 0.005 & 0.087 $\pm$ 0.009 & 26.35 / 46 & 0.835 $\pm$ 0.056 \\

\bottomrule
\end{tabular}
\centering\caption{Fit results at $\sqrt{s}$ = 2.76 TeV, using data from the ALICE collaboration~\cite{ALICE_2760}.}
\label{tab:results_2760}
\end{table}

\subsection{Fits to transverse momentum spectra for the interpolated data at 5.02 TeV}
For the transverse momentum spectra at $\sqrt{s}$ = 5.02 TeV we have used  the results  presented by the  
ALICE collaboration~\cite{ALICE_5020} (in a
 range extending up to about 20 GeV/c) by interpolating their own data at 2.76 TeV and 7 TeV. 
The fit is shown in Figure~\ref{pt_high}.  The resulting values of the  parameters are 
given in Table~\ref{tab:results_5020}. Again the values of $q$ 
and $T$ are much more constrained than for $\sqrt{s}$ = 900 GeV. 
As in the previous case, no common values of the parameters $T$ and $q$ can be found.
\begin{table}[!ht]
\centering
\begin{tabular}{|c|c|c|c|c|c|}\toprule
Particle       & $dN/dy$           & $q$                 & $T$ (GeV)   & $\chi_y^2$ / NDF & $\langle p_T \rangle$ (GeV)\\ \midrule

$\pi^++\pi^-$  & 4.452 $\pm$ 0.095 & 1.155 $\pm$ 0.002 & 0.076 $\pm$ 0.002 & 266.3 / 60 & 0.452 $\pm$ 0.016 \\

$K^++K^-$      & 0.530 $\pm$ 0.011 & 1.150 $\pm$ 0.005 & 0.099 $\pm$ 0.006 & 12.11 / 55  & 0.750 $\pm$ 0.049 \\


$p+\bar{p}$    & 0.235 $\pm$ 0.007 & 1.126 $\pm$ 0.005 & 0.091 $\pm$ 0.009 & 18.89 / 46 & 0.877 $\pm$ 0.059 \\

\bottomrule
\end{tabular}
\centering\caption{Fit results at $\sqrt{s}$ = 5.02 TeV, using data from the ALICE collaboration~\cite{ALICE_5020}.}
\label{tab:results_5020}
\end{table}

We would like to emphasize that all these results will have to be confirmed by the final analysis of the 5.02 TeV data.
\subsection{Fits to transverse momentum spectra at the beam energy of 7 TeV}

The transverse momentum spectra at $\sqrt{s}$ = 7 TeV in $p-p$ collisions have also been measured~\cite{ALICE_7000} 
in a range extending up to about 20 GeV/c. The fit is shown in Figure~\ref{pt_high} and the resulting values 
of the 
 parameters are given in Table~\ref{tab:results_7000}. 

\begin{table}[!ht]
\centering
\begin{tabular}{|c|c|c|c|c|c|}\toprule
Particle       & $dN/dy$           & $q$                 & $T$ (GeV)           & $\chi_y^2$ / NDF & $\langle p_T \rangle$ (GeV)\\ \midrule


$\pi^++\pi^-$  & 4.778 $\pm$ 0.101 & 1.158 $\pm$ 0.002 & 0.076 $\pm$ 0.002 & 331.7 / 55 & 0.460 $\pm$ 0.017 \\


$K^++K^-$      & 0.573 $\pm$ 0.011 & 1.155 $\pm$ 0.005 & 0.100 $\pm$ 0.006 & 27.54 / 48 & 0.777 $\pm$ 0.052 \\


$p+\bar{p}$    & 0.251 $\pm$ 0.007 & 1.129 $\pm$ 0.005 & 0.094 $\pm$ 0.009 & 20.26 / 46 & 0.903 $\pm$ 0.061 \\

\bottomrule
\end{tabular}
\caption{Fit results at $\sqrt{s}$ = 7 TeV, using data from the ALICE collaboration~\cite{ALICE_5020,ALICE_7000}.}
\label{tab:results_7000}
\end{table}

\section{Analysis of the  results}

\begin{figure}
\centering
\includegraphics[width=\textwidth, height=14.0cm]{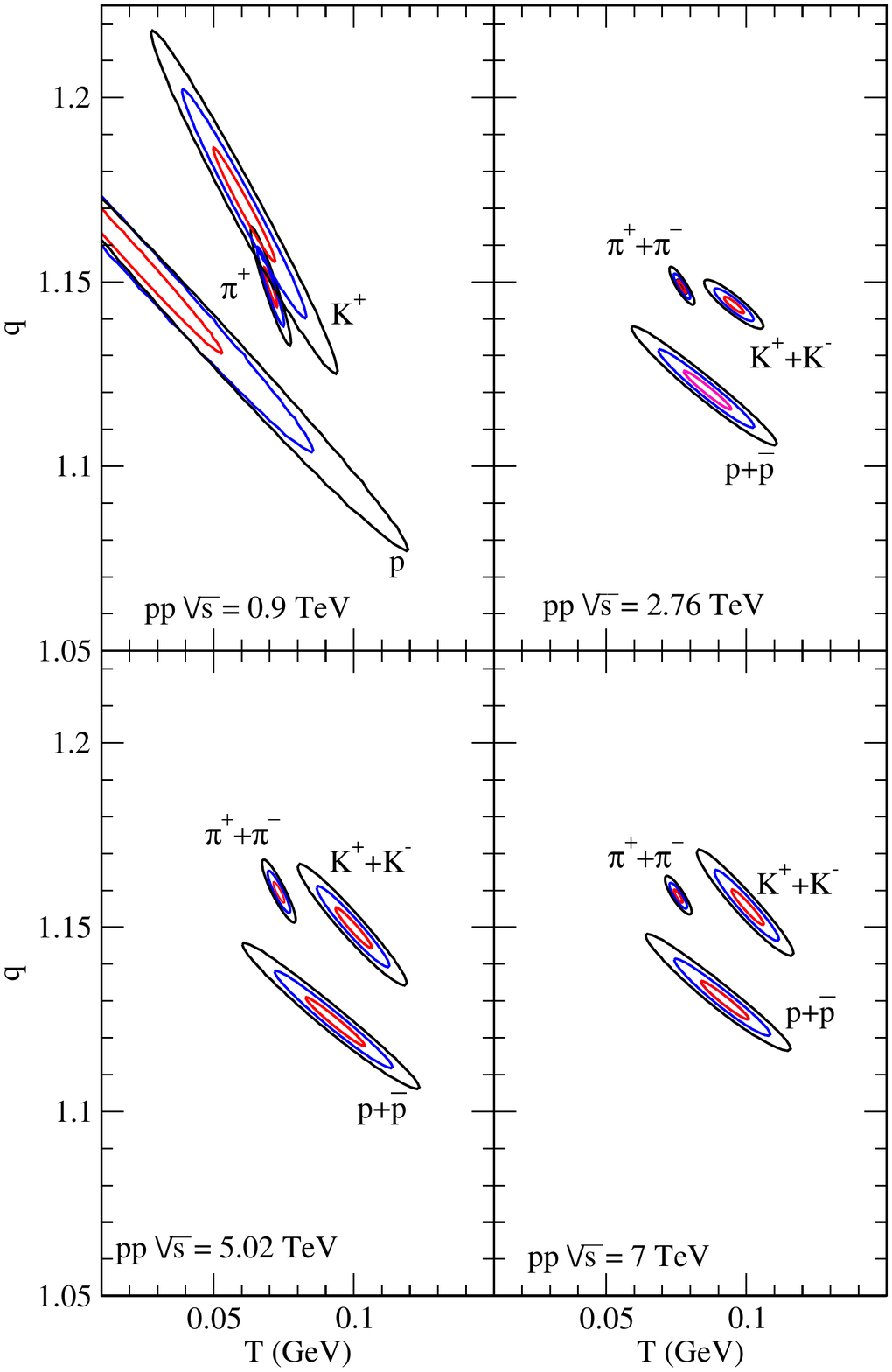}
\caption{\label{contours}
Contours in The $T - q$ plane showing lines of 1-$\sigma$ deviation from the minimum $\chi^2$
 in red. 2-$\sigma$ deviations are shown in blue. Those for 3-$\sigma$ deviation from the minimum $\chi^2$ are 
shown in black. The upper left pane is for data at $\sqrt{s}$ = 0.9 TeV~\cite{ALICE2}; the upper right pane is
for 2.76 TeV~\cite{ALICE_2760}. The lower left pane is for 5.02 TeV~\cite{ALICE_5020} while the lower 
right pane is for 7 TeV~\cite{ALICE_7000}. 
        }
\end{figure}

In Figure~\ref{contours} we show contours in the $T - q$ plane. The ellipses correspond to fixed values of deviations from the 
minimum $\chi^2$ values, 1-$\sigma$ uncertainties on the fit parameters are shown in red, 2-$\sigma$ uncertainties are shown 
in blue, while those with 3-$\sigma$ uncertainties are shown in black. 

At $\sqrt{s} = $ 0.9 TeV (upper left-hand pane in Figure~\ref{contours})
it can be seen that while the pions and kaons do overlap in a small region, 
this is not the case for the protons, albeit there is a wide range of possible values for the latter so that an eventual 
complete overlap for all three particles $\pi^+, K^+, p$ cannot really be excluded. A wider 
range in $p_T$ or a more precise measurement of the very low $p_T$ region is necessary to really exclude this eventuality.

At $2.76$ TeV (upper right-hand pane in Figure~\ref{contours}), it can be seen that the values 
of $q$ and $T$ are much more constrained  than in the 
previous case of 0.9 TeV. If the possibility of common values could not be discarded in the case of $0.9$ TeV, there is no doubt 
here that the values are different. 
The same goes for the $7$ TeV results (lower right-hand pane in Figure~\ref{contours}), and no common values of the parameters $T$ and $q$ can be 
found, nor any strict $m_T$ scaling because the parameters are different for each species.

We notice that the 7 TeV case has been discussed previously~\cite{wong,biro,azmi}, within the framework of the Tsallis 
distribution. The authors from~\cite{wong,biro} use a different form of the Tsallis distribution, not having a factor 
$m_T$ on the right-hand side of Eq.~(\ref{tsallisfit}) and hence they obtain higher values 
for the parameter $T$ (recalling that this is no longer a temperature in the thermodynamic sense).

 Within the framework considered here, the values for $q$ and $T$ 
obtained from the ALICE collaboration data are 
close to those of the CMS collaboration~\cite{CMS_7000}, as shown in the contour plot presented in Figure.~\ref{contours}. 
The CMS contours are indeed being situated as roughly equidistant from each of  the ALICE ones for pions, kaons and protons. 
This comes as a surprise since at those large values of the transverse momentum $p_T$, hard scattering processes are presumed to be 
dominant.

\begin{figure}
\centering
\includegraphics[width=\textwidth, height=14.0cm]{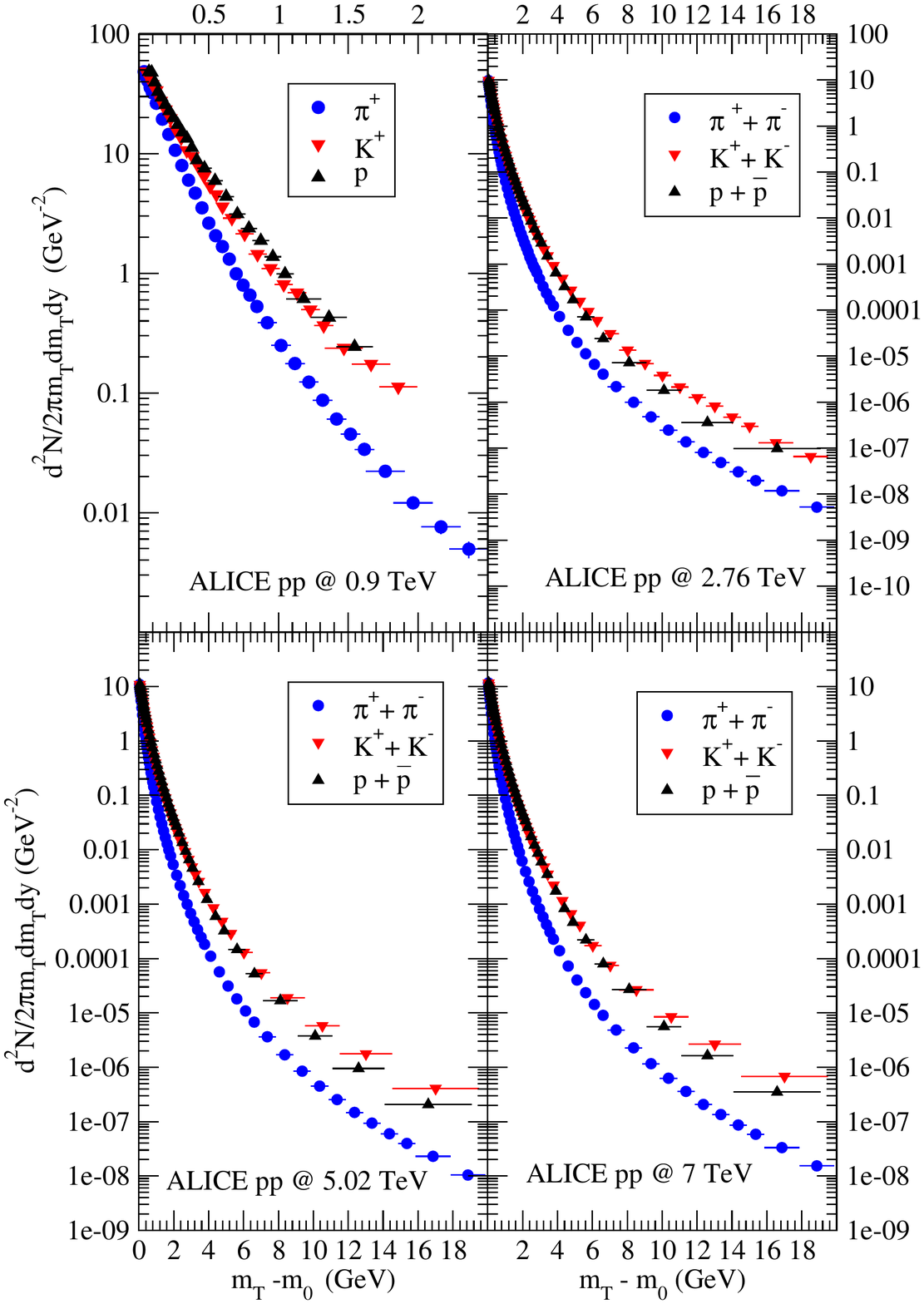}
\caption{\label{mT}
Transverse mass distributions for $\pi^+$, $K^+$ and protons at 900 GeV (top left)~\cite{ALICE2}. 
Also shown are $\pi^++\pi^-$, $K^++K^-$ and $p+\bar{p}$ at  2.76 TeV (top right)~\cite{ALICE_2760} 
5.02 TeV (bottom left)~\cite{ALICE_5020} and 7 TeV (bottom right)~\cite{ALICE_7000}. The data shown at
5.02 TeV are interpolated data.
        }
\end{figure}

In Figure~\ref{mT} we show the transverse mass distributions for pions, kaons and protons at 
different beam energies.  It is clearly seen that there is no $m_T$ scaling because of the  differences for each particle type. 

\begin{figure}
{\centering
\includegraphics[width=\columnwidth, height=15cm]{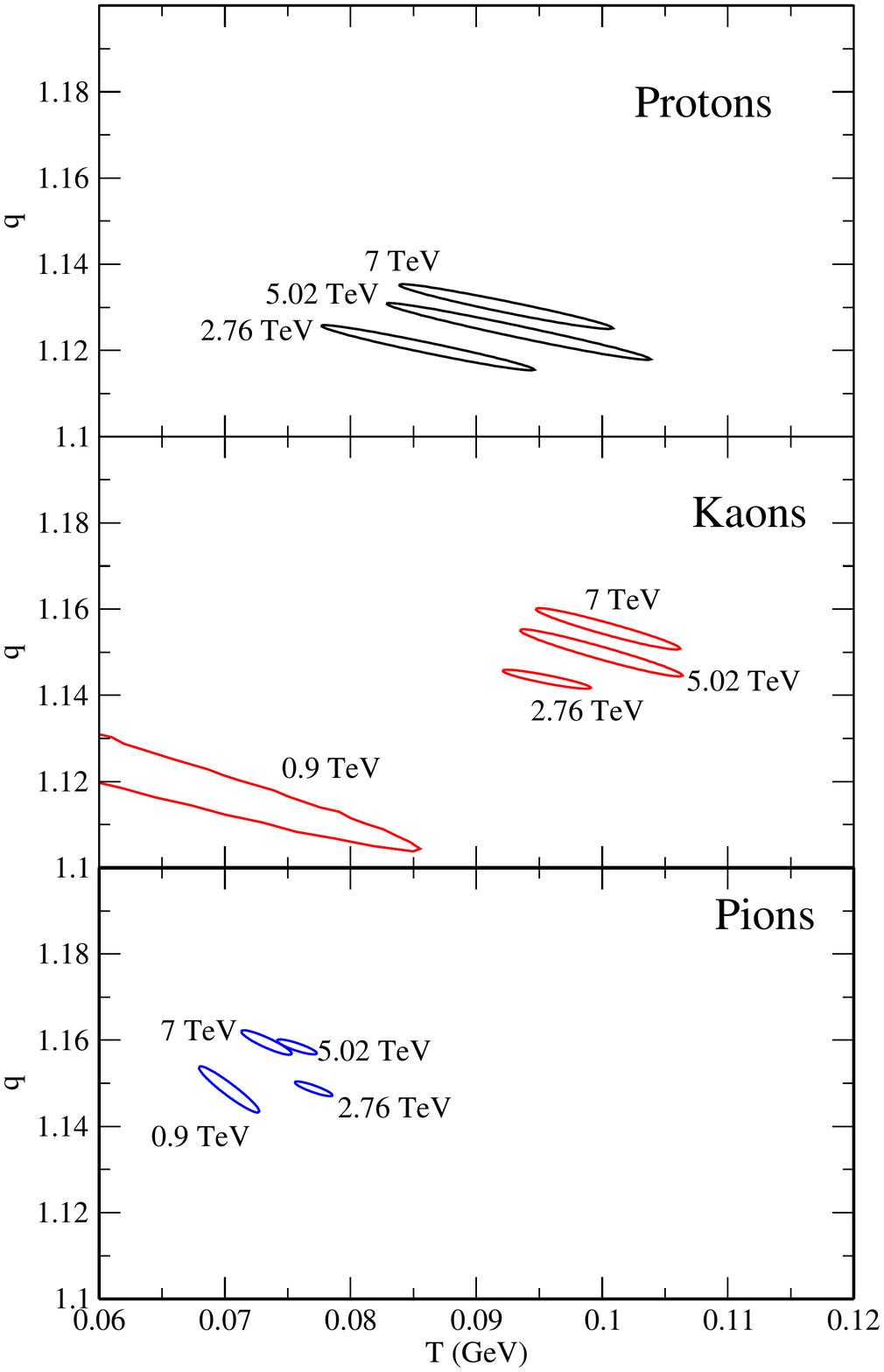}
\caption{Contours at fixed $\chi^2$ values corresponding to 1~--~$\sigma$ uncertainties, for protons (top), kaons (middle) and 
pions (bottom). Notice that for the 0.9 TeV contours, pions and kaons respectively mean $\pi^+$ and $K^+$, and that the 
0.9 TeV proton contour is simply out of range. The beam energies are displayed in the figure, and our results are 
obtained using results from the ALICE 
collaboration~\cite{ALICE2,ALICE_2760,ALICE_5020,ALICE_7000}.}
\label{contours_final}}
\end{figure}

In Figure~\ref{contours_final} we summarize some of the results separately for protons (upper pane),
kaons (middle pane) and pions (lower pane). 
We  notice here that our results clearly show that the fitted temperature $T$ is often comparable for kaons and 
for protons, which does not have a natural explanation in the radial flow scenario. In addition, the fitted $q$ parameter is 
also often comparable for pions and kaons. As a consequence, different groups of nearly $m_T$ scaling 
appear (see Figure~\ref{mT}), groups whose characteristic allowing for differentiating them could well be the mass 
range of the given particles.

We also notice that the  resulting values for $dN/dy$ at mid-rapidity are fully compatible (if not identical) to the values quoted by the 
ALICE collaboration using a slightly different parameterization for the transverse momentum distribution.
%
%

Finally there is also a clear beam energy dependence in the values of the parameters. 
The results obtained at 0.9 TeV are out of line, presumably because of the limited range in $p_T$ at this beam energy. 
Discarding for the moment the results at 0.9 TeV, one can see a clear shift for the pions towards  
higher values for $q$ but unchanged values of $T$. For protons the opposite result is seen, namely a shift towards higher temperatures 
but the values for $q$ are almost independent on the beam energy. For kaons the pattern is different again, namely an almost 
constant value for the temperature $T$ but a clear increase in the value of $q$.

\section{Large Transverse Momenta}
\begin{figure}
{\centering
\includegraphics[trim=0cm 0cm 0cm 0cm, clip=true,width=\columnwidth, height=18cm]{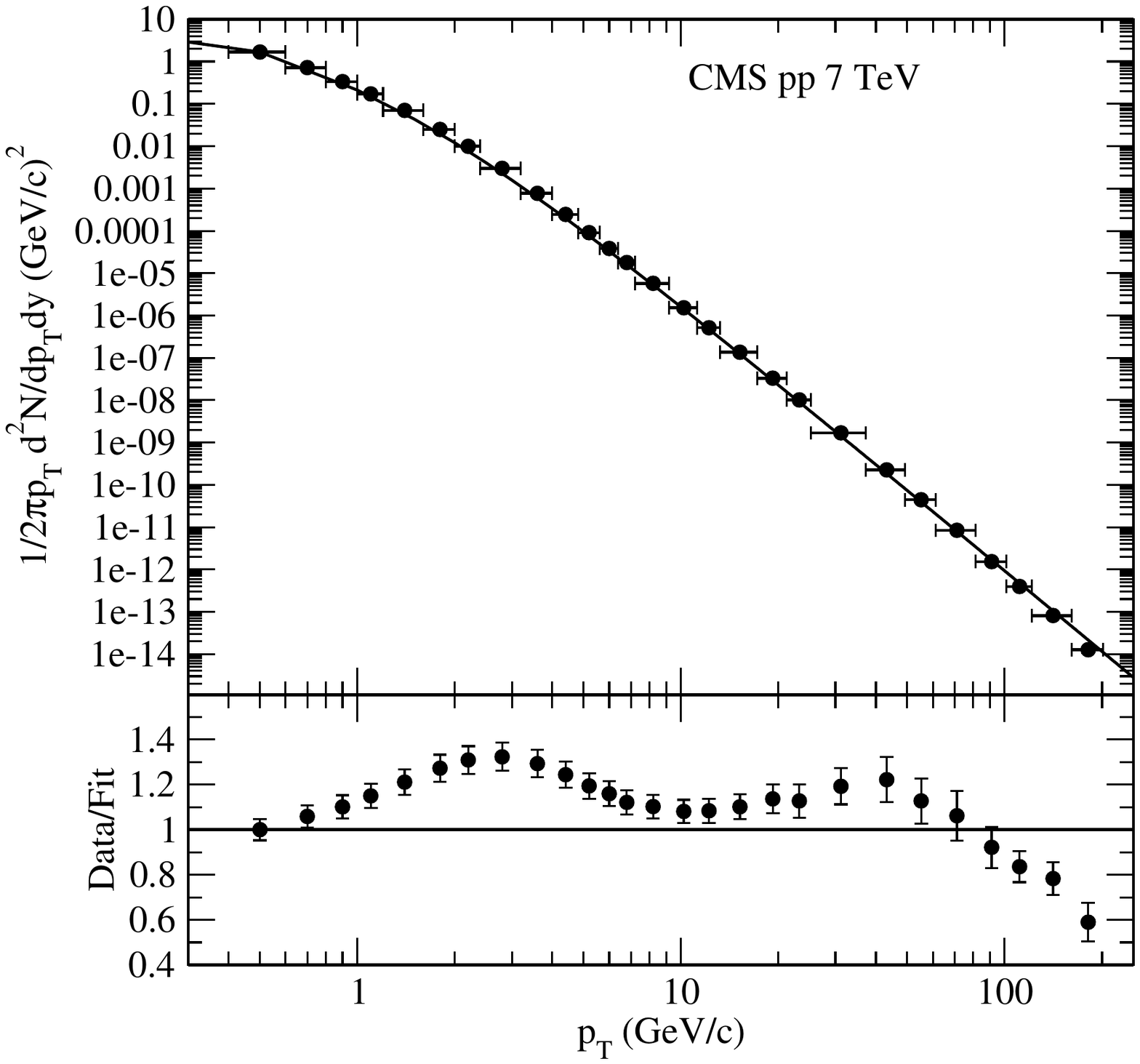}
\caption{
Results from the CMS collaboration~\cite{CMS_highpt} showing the transverse momentum distribution for charged particles 
up to 200 GeV/c. The fit is obtained by adding the contributions of $\pi^++\pi^-$, $K^++K^-$, $p + \bar{p}$
as obtained at a beam energy of 7 TeV in $p-p$ collisions using the Tsallis distributions with the parameters obtained 
from fitted ALICE data going only up to about 20 GeV/c as  
listed in Table 4.  The lower pane of the figure shows the ratio of data over the fit values. 
\label{highpt}
}}
\end{figure}

%
In order to get a feeling for the limit of applicability of the Tsallis distribution given in Eq.~\ref{tsallisfit} we applied the results obtained
above to the highest transverse momenta measured by the CMS collaboration~\cite{CMS_highpt} for charged particles up to 200 GeV/c in
$p-p$ collisions at a beam energy of 7 TeV.
To achieve this we simply 
 added  $\pi^++\pi^-$, $K^++K^-$, $p + \bar{p}$
using the parameters listed in Table 4.  
This has been done using the following expression:
\begin{equation}
\left.\frac{1}{2\pi p_T}\frac{d^{2}N_{ch}}{dp_T~dy}\right|_{y=0} = 
\frac{2}{(2\pi)^3}\sum_{i=1}^{3}\frac{g_im_{T,i}V_i}{(2\pi)^2}\left[1+(q_i-1)\frac{m_{T,i}}{T_i}\right]^{-\frac{q_i}{q_i-1}}
\end{equation}
where $i = \pi^+, K^+, p$. The relative weights between particles were determined by the 
corresponding degeneracy factors and given by $g_{\pi^+}$ = $g_{K^+}$ = 1 and $g_p$ = 2. 
The factor 2 on the right hand side shows the contribution from the antiparticles, $\pi^-, K^-$ and $\bar{p}$. \\
There are of course  more charged particles than 
just $\pi^+$, $K^+$ and protons (and their respective antiparticles) used here but it should be  close to a 
good approximation of the total number of charged particles.
The result is shown in Figure~\ref{highpt} and the description of the data is quite appropriate.
The lower pane of the figure shows the ratio of experimental data over fit values. There are indications that the
description using the Tsallis distribution deviates from measurements for higher values of $p_T$ but this is also the region where the uncertainty is largest.
It is to be concluded that the Tsallis distribution provides a reasonably good description up to the highest measured values of $p_T$.  
The oscillations 
seen in previous fits~\cite{Wilk:2014zka,Wilk:2015hla,Wilk:2017coo} are due to a simplified use of the Tsallis distribution which 
does not distinguish between
pions, kaons and protons.

\section{Conclusions}

In this paper we have thoroughly investigated one particular form of power tail distribution 
which is based on non-extensive statistical thermodynamics and has the property of having 
consistent thermodynamic relations for the particle number, the energy, the pressure and the entropy.

We have determined the parameters appearing in the Tsallis distribution as precisely as possible at beam 
energies of $\sqrt{s}$ = 0.9, 2.76 and 7 TeV. The conclusion we reach is that for the $\pi$'s, $K$'s 
and protons the parameters are different and no universality in the parameters exist. 
At the beam energy of 
$\sqrt{s}$ = 0.9 TeV, the interval in transverse momentum is fairly narrow and the uncertainty on the 
parameters is large so that the original analysis~\cite{worku1,worku2} made such a scenario possible.

The suggestion, made a few years ago, that the parameters appearing in this distribution are the same 
for a wide range of identified hadrons~\cite{worku1,worku2} at $\sqrt{s}$ = 900 GeV in p-p collisions 
is therefore  not supported by the analysis presented here.

Thus, even though the Tsallis distribution provides a reasonable description of the transverse momenta 
distributions, it has to be concluded that the parameters are clearly different. As a consequence, one 
basic property of the Tsallis distribution, namely scaling in the transverse mass $m_T$ is not obeyed 
because the relevant parameters change for different hadrons. 

At last, our conclusions are as follows:
\begin{itemize}
\item The Tsallis formula provides reasonably  good fits to the $p_T$ distributions in 
$p-p$ collisions at the LHC using three parameters $dN/dy$, $T$ and $q$.
\item The parameters $T$ and $q$ depend on the particle species and are different for pions, kaons and protons. No universal
behavior has been found.
\item As a consequence of this, there is no $m_T$ scaling.
\end{itemize}

\vspace{1cm}
\Large{{\bf Acknowledgments}}\\
\normalsize
T.~B.~would like to acknowledge the University Research Committee, University of Cape Town, South Africa for its 
financial support. S.~M.~would like to acknowledge the financial support from the Claude Leon Foundation.
The work of J.C. is based on research supported by the National Research Foundation (NRF) of South Africa.

\bibliographystyle{Science}
\bibliography{ReferencesFile}

\end{document}